\documentclass{easychair}

\usepackage[utf8]{inputenc}
\usepackage[T1]{fontenc}
\usepackage{graphicx}
\usepackage{url}

\title{A\"ira: Rethinking AI Research Assistants for Interdisciplinary Science}
\author{
Diya Mirji\inst{1}
\and
Tiffany Degbotse\inst{1}
\and
Sharmil Nanjappa\inst{1}
\and
Jiayi Zhou\inst{1,2}
\and
Vishnu Manoj\inst{1}
\and
Vihaan Nama\inst{1}
\and
Boyuan Chen\inst{1}
\and
Lee Tiedrich\inst{1,3}
\and
Christopher Bail\inst{4}
\and
David W. Johnston\inst{2}
\and
Walter Sinnott-Armstrong\inst{5}
\and
Brinnae Bent\inst{1}
}

\institute{
Pratt School of Engineering, Duke University, Durham, NC
\and
Nicholas School of the Environment, Duke University, Durham, NC
\and
Information Sciences School, University of Maryland, Baltimore, MD
\and
Department of Sociology, Duke University, Durham, NC
\and
Department of Philosophy and Kenan Institute for Ethics, Duke University, Durham, NC
}

\authorrunning{Mirji et al.}
\titlerunning{A\"ira: Rethinking AI Research Assistants}

\begin{document}

\maketitle

\begin{abstract}
Scientific discovery increasingly depends on interdisciplinary teams whose members contribute distinct expertise, conceptual frameworks, vocabularies, assumptions, and standards of evidence. Today's AI research assistants are largely designed to support individual researchers through literature review, writing assistance, coding, and data analysis. While these capabilities improve personal productivity, they provide little support for the collaborative reasoning required to integrate knowledge across disciplines. In this vision paper, we argue that AI research assistants should evolve from tools that optimize individual workflows to systems designed for interdisciplinary teams. We introduce a\"ira, an AI research assistant built around this idea. Rather than focusing solely on summarization or question answering, a\"ira identifies disciplinary perspectives, translates terminology, highlights assumptions, and synthesizes collaborative research opportunities. This paper offers a vision describing the design principles underlying a\"ira, presenting its system architecture, illustrating its outputs through interdisciplinary research meetings, and outlining future research directions for AI systems that support collaborative scholarship.
\end{abstract}

\section{Introduction}
\label{sec:introduction}

Scholarly inquiry increasingly relies on interdisciplinary collaboration \cite{wuchty2007increasing,evans2011metaknowledge,wu2019large}. Complex problems in health, climate, engineering, and public policy require researchers with different expertise to work together. However, interdisciplinary collaboration introduces challenges that extend beyond access to information: different fields often operate with different vocabularies, assumptions, methods, and conceptual frameworks, making it difficult for teams to integrate ideas and develop a shared understanding of research problems \cite{hall2012four,stokols2008science}.

At the same time, recent advances in large language models have transformed the way researchers interact with scientific information. AI research assistants now support activities such as literature retrieval, paper summarization, writing assistance, coding, and data analysis \cite{zhang2025exploring,zheng2025automation}. These systems have become valuable tools for accelerating individual research workflows, but most share a common orientation: they are built to support an individual researcher. The typical interaction consists of a single user asking questions, providing documents, or requesting assistance with a specific task. Current systems struggle to reliably bridge connections between disciplines and support human team coordination \cite{shen2026idrbench,cooper2026teamwork,super2026healthcare}. The
limitations of AI systems in team science reflect a fundamental gap in how the system is designed.

This individual-centered model is poorly matched to many of the challenges of interdisciplinary science. Productive collaboration depends not only on understanding the scientific problem itself, but also on understanding how collaborators approach that problem from different disciplinary perspectives. Researchers may use different terminology to describe similar concepts, rely on different methodological assumptions, apply different standards of evidence, and prioritize different research objectives. As a result, interdisciplinary teams need support for collaborative reasoning: defining problems together, reconciling competing perspectives, establishing shared terminology, identifying areas of agreement and disagreement, and translating discussion into actionable research plans.

In this vision paper, we introduce the argument that AI research assistants should evolve to support interdisciplinary collaboration. Rather than focusing exclusively on answering questions or generating content for individual users, future systems should help research teams develop shared understanding by organizing perspectives, translating terminology, highlighting assumptions, preserving unresolved questions, and structuring discussions into future research directions. These collaborative assistants can work alongside individual productivity tools in the research ecosystem, not competitively, but as tools for different purposes. We present a\"ira (\textbf{AI} \textbf{I}nterdisciplinary \textbf{R}esearch \textbf{A}ssistant), which illustrates how AI systems can be designed around the needs of research teams rather than individual researchers and serves as a step toward a new generation of AI assistants that facilitate collaborative scientific research.

\section{Design Principles for AI-Assisted Interdisciplinary Collaboration}
\label{sec:design-principles}

Current AI research assistants are primarily designed to retrieve information, answer questions, or generate text for individual users \cite{suli2025survey}. Supporting interdisciplinary teams requires a different set of capabilities. Based on the challenges encountered in interdisciplinary collaboration \cite{wuchty2007increasing,evans2011metaknowledge,wu2019large,fortunato2018science,hall2012four,stokols2008science}, we identified a set of design principles that guided the development of a\"ira and may serve as a foundation for future AI systems designed to support collaborative research.
\begin{enumerate}
\item First, AI systems should explicitly represent disciplinary perspectives rather than treating a discussion as a single, unified narrative. Researchers from different fields often approach the same problem with distinct objectives and standards of evidence. Preserving these perspectives allows teams to understand how each discipline contributes to the broader research effort.

\item Second, AI systems should translate terminology across disciplines. Many misunderstandings arise because the same term carries different meanings in different fields. Identifying these relationships can reduce communication barriers and establish a shared language.

\item Third, AI systems should highlight implicit assumptions. Research discussions frequently rely on unstated assumptions about methodology, data quality, causal relationships, or domain knowledge. Making these assumptions explicit allows collaborators to identify potential sources of disagreement and evaluate whether conclusions are based on shared premises.

\item Fourth, AI systems should distinguish between areas of agreement and disagreement rather than producing a single synthesized summary. Productive interdisciplinary collaboration depends on understanding these areas, and explicitly representing these points of alignment and divergence can help guide future discussion and decision-making.

\item Fifth, AI systems should preserve unresolved questions instead of forcing premature consensus. Many interdisciplinary meetings conclude with open questions that require additional expertise, data collection, or experimentation. Rather than treating these as incomplete discussions, AI systems should recognize them as important outcomes of the scientific process.

\item Finally, AI systems should actively support future collaboration. Beyond documenting what occurred during a meeting, they should organize discussions into actionable research artifacts that identify knowledge gaps, prioritize next steps, and develop new directions for investigation.

\end{enumerate}

\section{A\"ira: AI Research Assistant for Interdisciplinary Teams}
\label{sec:aira}

To explore the proposed design principles, we developed a\"ira, an AI research assistant designed to support interdisciplinary scientific collaboration. A\"ira supports interdisciplinary meetings by transforming recorded discussions into structured collaboration frameworks. Rather than generating a conventional meeting summary, a\"ira analyzes the discussion through the lens of interdisciplinary collaboration, identifying each discipline's perspectives, priorities, assumptions, and terminology before synthesizing these into a shared representation of the discussion. A\"ira does require researchers from multiple disciplines to attend the meeting and engage in meaningful discussion in order to synthesize a useful collaboration plan.

\begin{figure}[tb]
\centering
\includegraphics[width=0.7\textwidth]{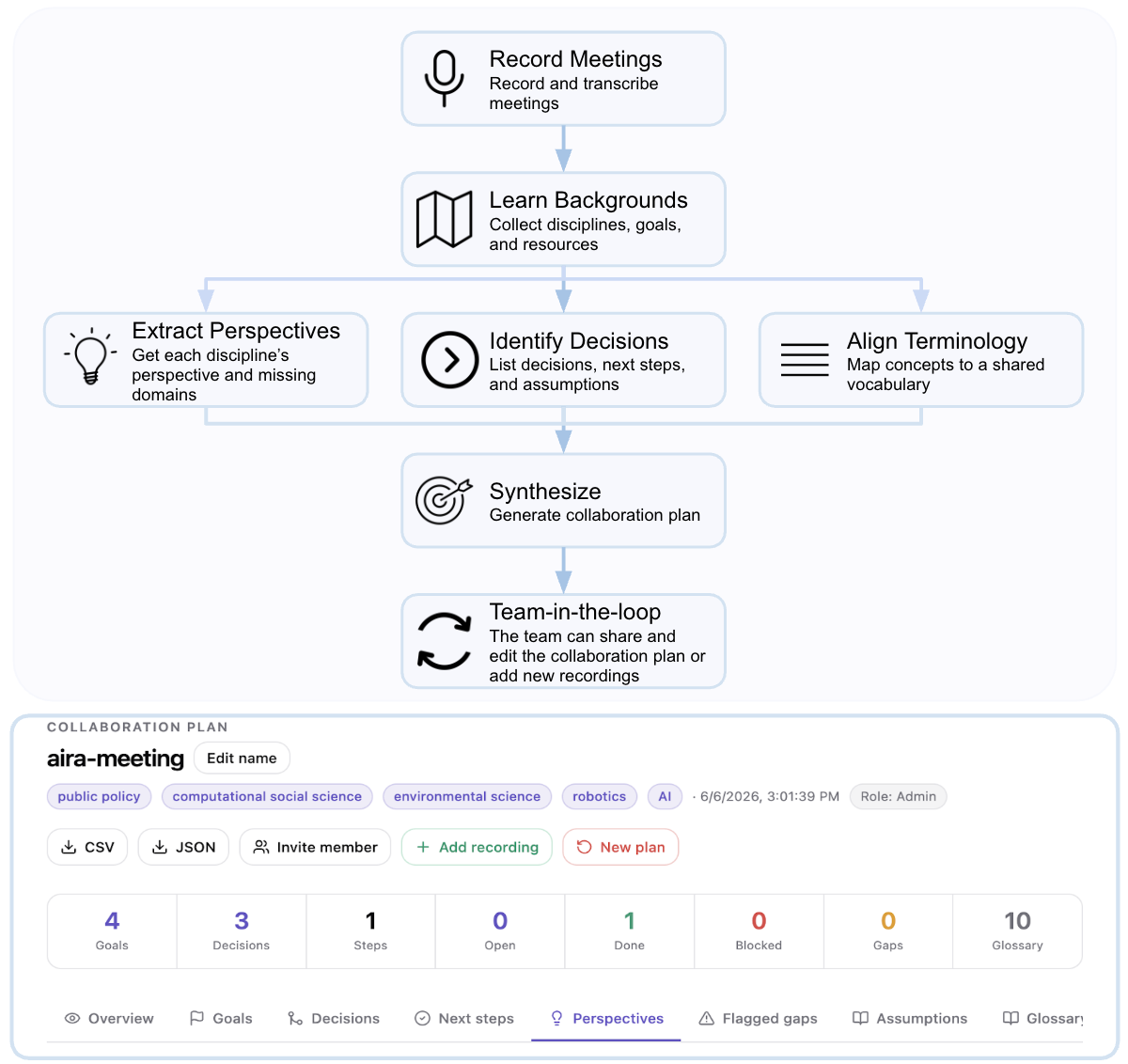}
\caption{Top: A\"ira workflow for generating a structured collaboration plan from interdisciplinary meeting transcripts. Bottom: Screenshot from a\"ira platform}
\label{fig:workflow}
\end{figure}

The overall workflow is illustrated in Figure~\ref{fig:workflow}. The process of creating a collaboration plan uses an agentic workflow of four agents. Following transcript ingestion, a\"ira's first agent extracts the primary disciplines represented in the discussion and collects the mentioned goals and resources. Then, with the extracted disciplines as context, the rest of the agents can run in parallel to gather different aspects of the collaboration plan. One agent derives the perspectives from each of the disciplines and then uses a fine-tuned LLM to highlight any missing disciplines that add a new viewpoint for future discussions. Another agent obtains the decisions that were discussed, including the rationale, dissent, and tradeoffs. It also lists the next steps with proposed actions and assumptions that the team made while making decisions. The last agent forms a glossary of terms and the discipline's definitions. Resulting insights are organized into a structured collaboration framework that can be shared and edited. Appendix~\ref{appendix} includes additional technical details. A\"ira can be found at [redacted for double blind submission].

We deployed a\"ira during an interdisciplinary summer research program involving philosophers and neuroscientists as a companion tool for research meetings involving philosophers and neuroscientists (IRB \#[redacted for double blind review]) (Appendix ~\ref{appendix}). This initial deployment was intended to explore the feasibility of the proposed workflow rather than to evaluate its effectiveness. Participants described the structured outputs as helpful for "clarifying unfamiliar terminology", "highlighting complementary expertise", and "organizing discussions into potential research directions". These qualitative observations motivate future studies evaluating AI systems designed to support interdisciplinary research.

\section{Future Directions}
\label{sec:future-directions}

Designing AI research assistants around interdisciplinary teams opens opportunities well beyond meeting support. As these systems mature, they could assist throughout the scientific lifecycle by helping research groups build shared understanding and maintain continuity across projects that span months or years. Possible applications include collaborative grant development, cross-disciplinary literature synthesis, and support for distributed research teams. Across these settings, future AI assistants could help teams align objectives and recognize conceptual connections that might otherwise remain isolated within disciplinary boundaries.

Furthermore, AI systems designed around collaborative reasoning may contribute more directly to scientific discovery. By integrating perspectives from multiple disciplines, tracking unresolved questions over time, and identifying gaps between existing knowledge and emerging hypotheses, future assistants may help research teams formulate new interdisciplinary research directions for investigation. These opportunities also raise important evaluation challenges. Current benchmarks for AI research assistants largely focus on individual tasks, while evaluation frameworks for collaboration-focused AI systems remain underdeveloped. Future work should investigate how to measure improvements in communication across disciplines, preservation of diverse perspectives, identification of conceptual gaps, and the quality of collaborative decision-making. This framework will guide formal evaluation of the efficacy of interdisciplinary AI assistants like a\"ira against baseline LLM-based workflows. We view a\"ira as an initial step toward this broader vision: a proof of concept for AI research assistants designed to support the collaborative processes of modern science.

\section{Conclusion}
\label{sec:conclusion}

As scientific problems become increasingly complex, effective collaboration across disciplines becomes increasingly important; and as research becomes increasingly interdisciplinary, AI research assistants should evolve beyond optimizing individual productivity. As a vision paper, we argue for a shift toward AI systems designed to support collaborative reasoning and present a\"ira as an initial step in that direction. We hope this work encourages the development of AI research assistants that strengthen communication, integration, and discovery across disciplines. 

\section*{Acknowledgements}
% Redacted for double blind submission.
This work was supported in part by OpenAI through the Duke DeepTech AI for Metascience Initiative.

\newpage
\appendix
\section{Appendix}
\label{appendix}
\subsection*{Technical Details}

\qquad A\"ira uses a multi-agent workflow of four main agents using OpenAI's gpt-4o model. All four agents get the same meeting transcript, but four different tasks. The workflow starts with the user entering the meeting by providing a transcript of the meeting, providing a video or audio recording of the meeting, or directly recording the meeting in the browser. If a recording is given, OpenAI's Whisper model is used for speech-to-text transcription. 

This transcription is passed to the first agent. The first agent is prompted to act like a research analyst extracting structured information from an interdisciplinary research meeting transcript. The resulting JSON returns a unique plan id to name the plan, academic and professional disciples, stated or implied goals with their priorities, and mentioned roles, tools, and datasets as resources. We ask the agent to be conservative and only extract what is clearly present in the transcript.

Once the disciplines are extracted from the first agent, the remaining three can run in parallel. The second agent is prompted to act as an interdisciplinary research facilitator specializing in perspective synthesis. From the meeting transcript and the identified disciplines, the agent is asked to learn how each discipline frames the research problem and to identify missing voices whose absence could complement the research. Each domain flagged as a gap includes an impact measure (high, medium, or low) and an explanation citing a specific issue that this domain improves the perspective. There is a large margin for hallucination on suggesting missing perspectives, so we provide a fine-tuned LLM that adds enriched perspectives. The LLM is a gpt-4o-mini model that is fine-tuned on synthetic conversations and their missing perspective labels. The data consisted of 50 synthetic transcripts between neuroscientists and philosophers and the missing perspectives from each conversation as the label, which was grounded in academic documents sourced from professors in the fields of philosophy, aerial systems, neuroscience, environmental science, and robotics. The training examples were used to fine-tune the gpt-4o-mini with validation loss evaluation across three epochs. This fine-tuned LLM will propose a domain that might be missing and its rationale so that the second agent can decide whether the suggested missing domains were reasonable in the context of the present discipline's perspectives. To focus the agent, we require skepticism with only essential domains that relate to specific sentences in the transcript and are traceable through reasoning.

The third agent is prompted to act as an expert at parsing research meeting discourse. It is asked to map disagreements between disciplines along with what was gained or lost, the affected domains, and the severity of the disagreement. Additionally, it extracts suggested next steps, uncertainty or "if...then" conditions as assumptions, agreements as decisions with their rationale. We instruct this agent not to invent content that is not present in the transcript. 

The fourth agent is prompted to act as a cross-disciplinary terminology specialist that identifies domain-specific terms in research meeting transcripts and builds a shared glossary. It is instructed to focus on terms that are likely to cause confusion across disciplines along with their definition. The related terms for each identified term are concepts from other disciplines that refer to similar or overlapping topics. 

\subsection*{Data Privacy}

\qquad A\"ira has a privacy policy that alerts users of the data that are stored. When signing in with the user's Google account, we collect the name, email address, and profile picture if provided. When using a\"ira to generate a collaboration plan, the user can provide a meeting transcript (pasted text, uploaded file, or live recording). This input is processed to generate a plan. We do not store the raw transcript or audio after the plan is generated. The collaboration plans generated from the input, edits made to the plans, and project membership information if collaborators are invited are stored. A\"ira does not sell user personal information. 

A\"ira's third-party services are Google OAuth for signing in and authentication, OpenAI for transcription and plan generation, Supabase for database storage and authentication, and Railway for application hosting. The resulting collaboration plan is stored on the secured platform, Supabase, and configured with row-level security. Data is securely transmitted to models provided through the OpenAI API. Per OpenAI's API policies, submitted data is not used to train OpenAI's models. Lastly, recordings are not retained or used to train A\"ira. Unless we obtain the user's unformed consent for research purposes, we do not use any data for system improvement. 

\subsection*{IRB Study}
\qquad The goal of the study was to understand how AI-assisted synthesis of research discussions may improve shared understanding, integration of ideas, and the overall quality of collaborative research planning across disciplines. The study only collected voluntary pre-survey and post-survey responses to understand the feasibility of the workflow. We asked participants to complete a pre-survey before using a\"ira to collect consent and understand background information such as prior interdisciplinary research experience, confidence in cross-disciplinary communication, and initial perspectives on the research topic. The post-survey evaluates the participant’s collaboration experience, including perceived interdisciplinary integration, satisfaction with the collaboration, self-assessed proposal quality, time required to complete the task, and perceived usefulness of the AI tool. During this pilot study, the platform was deployed to 17 participants.

\end{document}